\documentclass[aps,prl,twocolumn,superscriptaddress,10pt,longbibliography]{revtex4-1}

\usepackage{amsmath}
\usepackage{ulem}
\usepackage{braket}

\usepackage{bm} % Bold math symbols in latex plots

\usepackage{color}

\usepackage{graphicx}
\usepackage{epstopdf}

\graphicspath{
{figures/}
}
\newcommand{\wminus}{\omega_{-}}
\newcommand{\wplus}{\omega_{+}}
\newcommand{\wfsr}{\bar{\omega}}

\newcommand{\stateAR}{\ket{\psi_{AR}}}
\newcommand{\stateR}{\ket{\psi_{R}}}
\newcommand{\stateARdelayed}{\ket{\overline{\psi}_{AR}}}
\newcommand{\stateRdelayed}{\ket{\overline{\psi}_{R}}}

\begingroup\lccode`~=`_
\lowercase{\endgroup\def~}#1{_{\scriptscriptstyle#1}}

\AtBeginDocument{\mathcode`_="8000 \catcode`_=12 }

\begin{document}

\title{Generation and symmetry control of high-dimensional quantum frequency states}
\author{G. Maltese}
\affiliation{Laboratoire Mat\'eriaux et Ph\'enom\`enes Quantiques, Universit\'e Paris Diderot, CNRS-UMR 7162, Paris 75013, France}
\author{M.I. Amanti*}
\affiliation{Laboratoire Mat\'eriaux et Ph\'enom\`enes Quantiques, Universit\'e Paris Diderot, CNRS-UMR 7162, Paris 75013, France}
\author{F. Appas}
\affiliation{Laboratoire Mat\'eriaux et Ph\'enom\`enes Quantiques, Universit\'e Paris Diderot, CNRS-UMR 7162, Paris 75013, France}
\author{G. Sinnl}
\affiliation{Laboratoire Mat\'eriaux et Ph\'enom\`enes Quantiques, Universit\'e Paris Diderot, CNRS-UMR 7162, Paris 75013, France}
\author{A. Lema\^itre}
\affiliation{Centre de Nanosciences et de Nanotechnologies, CNRS, Universit\'e Paris-Sud, Universit\'e Paris-Saclay, C2N-Marcoussis}
\author{P. Milman}
\affiliation{Laboratoire Mat\'eriaux et Ph\'enom\`enes Quantiques, Universit\'e Paris Diderot, CNRS-UMR 7162, Paris 75013, France}
\author{F.Baboux}
\affiliation{Laboratoire Mat\'eriaux et Ph\'enom\`enes Quantiques, Universit\'e Paris Diderot, CNRS-UMR 7162, Paris 75013, France}
\author{S. Ducci}
\affiliation{Laboratoire Mat\'eriaux et Ph\'enom\`enes Quantiques, Universit\'e Paris Diderot, CNRS-UMR 7162, Paris 75013, France}

\begin{abstract}
High-dimensional quantum states are promising resources for quantum communication and processing. In this context the frequency degree of freedom of light combines the advantages of robustness and easy handling with standard classical telecommunication components. In this work we propose a method to generate and control the symmetry of broadband biphoton frequency states, based on the interplay of cavity effects and relative temporal delay between the two photons of each pair. We demonstrate it using an integrated AlGaAs semiconductor platform producing quantum frequency combs, working at room temperature and compliant with electrical injection. These results open interesting perspectives for the development of massively parallel and reconfigurable systems for complex quantum operations.

\end{abstract}
\pacs{}
\maketitle
%%%%%%%%%%%%%%%%%%%%%%%%%%%%%%%%%%%%%%%%%%%%%%%%%%%%%%%%%%%%%%%%%%%%%%%%%%%%

Since the emergence of the domain of quantum information, quantum optics plays an important role as an experimental test bench for a large variety of novel concepts; nowadays, in the framework of the development of quantum technologies, photonics represents a promising platform for several applications ranging from long distance quantum communications to the simulation of complex phenomena and metrology \cite{Flamini_2018, giovannetti2011advances}. In these last years a growing attention has been devoted to large scale entangled quantum states of light as key elements to increase the data capacity and robustness in quantum information protocols. Such states can be realized through qubits encoded in many-particles, but this approach suffers from scalability problems; an alternative strategy is to work with a lesser number of particles and to encode information in high-dimensional states. This has been implemented using different degrees of freedom of light: spatial or path modes \cite{pathHD,PhysRevA.87.012326}, orbital angular momentum \cite{McLaren:12,zhang2016engineering}, time-energy \cite{timeenergyHD}, frequency  \cite{Kues,Imany18}. Among all these possibilities the frequency domain is particularly appealing thanks to its compatibility with the existing fibered telecom network \cite{frequencyBIN}; moreover, it enables the development  of robust and scalable systems in a single spatial mode, without the requirement of complex beam shaping or stabilized interferometers.

The most straightforward physical process to generate quantum states in the frequency domain is nonlinear optical conversion, widely used to produce photon pairs for quantum information and communications protocols. 

A convenient way to handle the frequency continuous degree of freedom is to discretize it and generate biphoton frequency combs \cite{kues2019quantum}. 
Such states have first been investigated exploiting spontaneous parametric down-conversion (SPDC) in dielectric crystals \cite{lu2003mode,Harnessing,firstpaperQdit}, by placing a resonant cavity either after  or around the nonlinear material. In the latter case the state is shaped directly at the generation stage with the advantage of avoiding signal reduction \cite{JMoreno2010}. More recently biphoton frequency combs have been generated in integrated optical micro-resonators via spontaneous four-wave mixing: this approach overcomes the drawbacks of low scalability and high cost of bulk systems. Interesting results on the generation and coherent manipulation of high-dimensional frequency states have been obtained in both Hydex \cite{Kues} and silicon nitride micro-rings \cite{Imany18}. 
For the development of accessible processing of quantum frequency combs a handy control over their symmetry is desirable. The ability to switch from symmetric to anti-symmetric high-dimensional states opens the way to the implementation of qudits teleportation, logic gates as well as dense coding and state discrimination \cite{Goyal2014,applcationmultipart,zhang2016engineering}. These concepts have started to be explored in \cite{Sagioro2004, olindo2006hong} making use of bulk Fabry-Perot cavities in Hong-Ou-Mandel (HOM) interferometry experiments.

In this work, we propose a method to generate and control the symmetry of biphoton frequency combs by combining the spectral filtering effect of a cavity with the control of the temporal delay between photons of a pair. We show that the simple tuning of the pump frequency allows to engineer the wavefunction symmetry. The advantage of our proposal is that it doesn't rely on post-selection, as usual techniques based on coincidence measurements of photons emerging from a beam-splitter. We demonstrate our method on an integrated AlGaAs semiconductor device emitting broadband frequency quantum states in the telecom range, working at room temperature and compliant with electrical injection \cite{PhysRevLett.112.183901}.
\begin{figure} [h]
\includegraphics[width=\columnwidth]{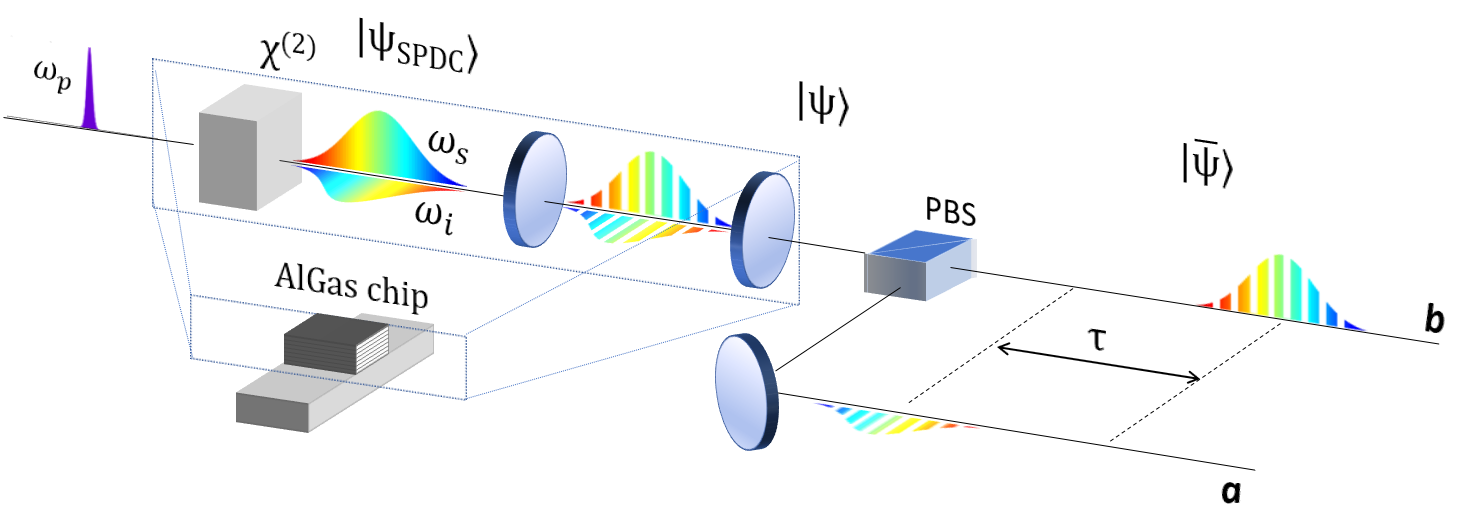}
\centering
\caption{Schematic of the experimental set-up for the generation and manipulation of biphoton frequency comb. A monochromatic pump beam generates photon pairs in the state $\ket{\psi_{SPDC}}$ by type II SPDC. An optical cavity discretizes the spectrum of the emitted photons, producing a biphoton combs $\ket{\psi}$. The photons of each pair are deterministically separated with a polarizing beam splitter (PBS) and an optical delay $\tau$ is imposed between them, leading to the state $\overline{\ket{\psi}}$. The symmetry of $\overline{\ket{\psi}}$ is controlled by tuning the pump frequency.}
\label{figsetup1}
\end{figure}
In Figure \ref{figsetup1} we present the schematic of the experimental set up for the generation and symmetry manipulation of biphoton frequency combs.
Photon pairs are generated by type II SPDC in a nonlinear medium: a pump photon at the frequency $\omega_p$ annihilates, generating two orthogonally polarized photons called  signal (at frequency $\omega_s$) and idler (at frequency $\omega_i$). The resulting quantum state can be written as: 
\begin{equation}
\begin{aligned}
\ket{\psi_{SPDC}}=\int_{-\infty}^{\infty} \int_{-\infty}^{\infty} C(\omega_s,\omega_i)\ket{H,\omega_s}\ket{V,\omega_i}d\omega_s d \omega_i
\end{aligned}
\end{equation}
where $C (\omega_s,\omega_i)$  is the joint spectral amplitude (JSA), i.e. the probability amplitude of having one of the photons at frequency $\omega_s$ with polarization $H$ and the other at frequency $\omega_i$ with polarization $V$.
For convenience we write the state using the basis $\wplus= \omega_s+\omega_i$ and $\wminus= \omega_s-\omega_i$; in this case, the JSA function takes the expression $C (\wplus,\wminus)=C_{p} (\wplus) C_{PM}(\wplus,\wminus)$, where $C_{p}$ is the pump spectral profile and $C_{PM}$ is the phase matching function depending on the material properties. For a pump frequency $\omega_p$ close to the degeneracy, $C_{PM}$ is in good approximation a symmetric function in $\wminus$, centered in $\wminus=0$, whose bandwidth depends on the characteristics of the nonlinear medium \cite{barbieri2017hong}.
An optical cavity discretizes the frequency space of the state  $\ket{\psi_{SPDC}}$ (See Figure \ref{figsetup1}). At this stage the function $C_{cav}$ associated to the cavity, which is the product of the signal and idler cavity transmission functions ($C_{cav}=T_s(\omega_s)T_i(\omega_i)=C_{cav}(\wplus,\wminus)$), modulates the state JSA. The resulting state is a biphoton frequency comb, consisting in a sequence of phase-locked evenly-spaced peaks with a common phase originating from the pump \cite{lu2003mode}. Figure \ref{JSI_Theory} (a) presents the numerical simulation of the corresponding joint spectral intensity (JSI=$|(C_{p} (\wplus) C_{PM}(\wplus,\wminus) C_{cav}(\wplus,\wminus)|^{2}$), which is the accessible function for experimental measurements. We show a zoom of the JSI around degeneracy, for a cavity consisting of a Fabry-Perot resonator of mirror reflectivity R=0.8, free spectral range $ \bar{\omega} $ and a pump laser of bandwidth $\Delta\omega \gg \bar{\omega} $.
We observe that the JSI presents a periodic pattern with a fixed periodicity $2\bar{\omega}$ in both $\wplus$ and $\wminus$ directions. 
The number of peaks in the $ \wplus$ direction is determined by the width of the pump spectral profile, while the one in the $\wminus$ direction is determined by the width of the $C_{PM}$ function. 

In the case of a monochromatic pump beam (linewidth $\Delta\omega \ll \bar{\omega} $), the pump spectral profile can be approximated as $C_{p} (\wplus)$ = $\delta(\wplus-\omega_p)$.  The corresponding quantum state at the output of the cavity is:

\begin{equation}
\begin{aligned}
\begin{split}
\ket{\psi}= \int_{-\infty}^{\infty} C_{PM}(\omega_p,\wminus) C_{cav}(\omega_p,\wminus) d \wminus\\
\ket{H,\frac{\omega_p + \wminus}{2}}\ket{V,\frac{\omega_p-\wminus}{2}}
\end{split}
\end{aligned}
\end{equation}
By tuning of the pump frequency we have access to two classes of states, having different spectral pattern (See Figure \ref{JSI_Theory}(a)). 
For $\omega_p=\omega_R= 2 n \bar{\omega}$, with $n$ integer number, we generate resonant states, whose JSI maxima are disposed at even multiple of $\bar{\omega}$ (See Figure \ref{JSI_Theory}(b)). In the approximation of a cavity with perfect reflectivity, the resonant state is:
\begin{equation}
\begin{aligned}
\begin{split}
\ket{\psi_R}= \sum_m \int_{-\infty}^{\infty} C_{PM}(\omega_R,\wminus) \delta(\wminus-2m\bar{\omega}) d \wminus\\
\ket{H,\frac{\omega_R + \wminus}{2}}\ket{V,\frac{\omega_R-\wminus}{2}}
\end{split}
\end{aligned}
\end{equation}
For $\omega_p=\omega_{AR}= (2n+1) \bar{\omega}$, we generate anti-resonant states, whose JSI maxima are disposed at odd multiples of $\bar{\omega}$ (See Figure \ref{JSI_Theory}(c)). In the same approximation of a cavity with perfect reflectivity the anti-resonant state is:
\begin{equation}
\begin{aligned}
\begin{split}
\ket{\psi_{AR}}= \sum_m \int_{-\infty}^{\infty} C_{PM}(\omega_{AR},\wminus) \delta(\wminus-(2m+1)\bar{\omega}) d \wminus\\
\ket{H,\frac{\omega_{AR} + \wminus}{2}}\ket{V,\frac{\omega_{AR}-\wminus}{2}}
\end{split}
\end{aligned}
\end{equation}

\begin{figure*}[ht]
% GNUPLOT: LaTeX picture with Postscript
\begingroup
\newcommand{\hl}[1]{\setlength{\fboxsep}{0.75pt}\colorbox{white}{#1}}
  \makeatletter
  \providecommand\color[2][]{%
    \GenericError{(gnuplot) \space\space\space\@spaces}{%
      Package color not loaded in conjunction with
      terminal option `colourtext'%
    }{See the gnuplot documentation for explanation.%
    }{Either use 'blacktext' in gnuplot or load the package
      color.sty in LaTeX.}%
    \renewcommand\color[2][]{}%
  }%
  \providecommand\includegraphics[2][]{%
    \GenericError{(gnuplot) \space\space\space\@spaces}{%
      Package graphicx or graphics not loaded%
    }{See the gnuplot documentation for explanation.%
    }{The gnuplot epslatex terminal needs graphicx.sty or graphics.sty.}%
    \renewcommand\includegraphics[2][]{}%
  }%
  \providecommand\rotatebox[2]{#2}%
  \@ifundefined{ifGPcolor}{%
    \newif\ifGPcolor
    \GPcolortrue
  }{}%
  \@ifundefined{ifGPblacktext}{%
    \newif\ifGPblacktext
    \GPblacktexttrue
  }{}%
  % define a \g@addto@macro without @ in the name:
  \let\gplgaddtomacro\g@addto@macro
  % define empty templates for all commands taking text:
  \gdef\gplbacktext{}%
  \gdef\gplfronttext{}%
  \makeatother
  \ifGPblacktext
    % no textcolor at all
    \def\colorrgb#1{}%
    \def\colorgray#1{}%
  \else
    % gray or color?
    \ifGPcolor
      \def\colorrgb#1{\color[rgb]{#1}}%
      \def\colorgray#1{\color[gray]{#1}}%
      \expandafter\def\csname LTw\endcsname{\color{white}}%
      \expandafter\def\csname LTb\endcsname{\color{black}}%
      \expandafter\def\csname LTa\endcsname{\color{black}}%
      \expandafter\def\csname LT0\endcsname{\color[rgb]{1,0,0}}%
      \expandafter\def\csname LT1\endcsname{\color[rgb]{0,1,0}}%
      \expandafter\def\csname LT2\endcsname{\color[rgb]{0,0,1}}%
      \expandafter\def\csname LT3\endcsname{\color[rgb]{1,0,1}}%
      \expandafter\def\csname LT4\endcsname{\color[rgb]{0,1,1}}%
      \expandafter\def\csname LT5\endcsname{\color[rgb]{1,1,0}}%
      \expandafter\def\csname LT6\endcsname{\color[rgb]{0,0,0}}%
      \expandafter\def\csname LT7\endcsname{\color[rgb]{1,0.3,0}}%
      \expandafter\def\csname LT8\endcsname{\color[rgb]{0.5,0.5,0.5}}%
    \else
      % gray
      \def\colorrgb#1{\color{black}}%
      \def\colorgray#1{\color[gray]{#1}}%
      \expandafter\def\csname LTw\endcsname{\color{white}}%
      \expandafter\def\csname LTb\endcsname{\color{black}}%
      \expandafter\def\csname LTa\endcsname{\color{black}}%
      \expandafter\def\csname LT0\endcsname{\color{black}}%
      \expandafter\def\csname LT1\endcsname{\color{black}}%
      \expandafter\def\csname LT2\endcsname{\color{black}}%
      \expandafter\def\csname LT3\endcsname{\color{black}}%
      \expandafter\def\csname LT4\endcsname{\color{black}}%
      \expandafter\def\csname LT5\endcsname{\color{black}}%
      \expandafter\def\csname LT6\endcsname{\color{black}}%
      \expandafter\def\csname LT7\endcsname{\color{black}}%
      \expandafter\def\csname LT8\endcsname{\color{black}}%
    \fi
  \fi
    \setlength{\unitlength}{0.0500bp}%
    \ifx\gptboxheight\undefined%
      \newlength{\gptboxheight}%
      \newlength{\gptboxwidth}%
      \newsavebox{\gptboxtext}%
    \fi%
    \setlength{\fboxrule}{0.5pt}%
    \setlength{\fboxsep}{1pt}%
\begin{picture}(9740.00,3740.00)%
    \gplgaddtomacro\gplbacktext{%
      \csname LTb\endcsname%
      \put(2435,3672){\makebox(0,0){\strut{}JSI}}%
    }%
    \gplgaddtomacro\gplfronttext{%
      \csname LTb\endcsname%
      \put(1423,374){\makebox(0,0){\strut{}-4}}%
      \csname LTb\endcsname%
      \put(1929,374){\makebox(0,0){\strut{}-2}}%
      \csname LTb\endcsname%
      \put(2435,374){\makebox(0,0){\strut{}0}}%
      \csname LTb\endcsname%
      \put(2941,374){\makebox(0,0){\strut{}2}}%
      \csname LTb\endcsname%
      \put(3447,374){\makebox(0,0){\strut{}4}}%
      \csname LTb\endcsname%
      \put(2435,115){\makebox(0,0){\strut{}$\wminus / \wfsr$}}%
      \csname LTb\endcsname%
      \put(1001,605){\makebox(0,0)[r]{\strut{}-3}}%
      \csname LTb\endcsname%
      \put(1001,1111){\makebox(0,0)[r]{\strut{}-2}}%
      \csname LTb\endcsname%
      \put(1001,1617){\makebox(0,0)[r]{\strut{}-1}}%
      \csname LTb\endcsname%
      \put(1001,2123){\makebox(0,0)[r]{\strut{}0}}%
      \csname LTb\endcsname%
      \put(1001,2629){\makebox(0,0)[r]{\strut{}1}}%
      \csname LTb\endcsname%
      \put(1001,3135){\makebox(0,0)[r]{\strut{}2}}%
      \csname LTb\endcsname%
      \put(720,1870){\rotatebox{-270}{\makebox(0,0){\strut{}$(\wplus - \omega_0) / \wfsr$}}}%
      \csname LTb\endcsname%
      \put(1170,3458){\makebox(0,0){\strut{}0}}%
      \csname LTb\endcsname%
      \put(3700,3458){\makebox(0,0){\strut{}1}}%
      \colorrgb{0.58,0.00,0.83}%
      \put(664,3236){\makebox(0,0){\strut{}\hl{\small{\textbf{(a)}}}}}%
      \csname LTb\endcsname%
      \put(3320,2274){\makebox(0,0){\strut{}\textcolor{white}{resonant}}}%
      \csname LTb\endcsname%
      \put(3194,1415){\makebox(0,0){\strut{}\textcolor{white}{anti-resonant}}}%
      \colorrgb{0.58,0.00,0.83}%
      \put(664,3236){\makebox(0,0){\strut{}\hl{\small{\textbf{(a)}}}}}%
    }%
    \gplgaddtomacro\gplbacktext{%
      \csname LTb\endcsname%
      \put(4412,2119){\makebox(0,0){\strut{}}}%
      \csname LTb\endcsname%
      \put(4860,2119){\makebox(0,0){\strut{}}}%
      \csname LTb\endcsname%
      \put(5308,2119){\makebox(0,0){\strut{}}}%
      \csname LTb\endcsname%
      \put(5755,2119){\makebox(0,0){\strut{}}}%
      \csname LTb\endcsname%
      \put(6203,2119){\makebox(0,0){\strut{}}}%
    }%
    \gplgaddtomacro\gplfronttext{%
      \csname LTb\endcsname%
      \put(4155,2803){\rotatebox{-270}{\makebox(0,0){\strut{}}}}%
      \csname LTb\endcsname%
      \put(5307,1940){\makebox(0,0){\strut{}}}%
      \colorrgb{0.06,0.35,0.80}%
      \put(4009,3270){\makebox(0,0){\strut{}\hl{\small{\textbf{(b)}}}}}%
      \colorrgb{0.06,0.35,0.80}%
      \put(5308,3503){\makebox(0,0){\strut{}$\stateR$}}%
    }%
    \gplgaddtomacro\gplbacktext{%
      \csname LTb\endcsname%
      \put(4412,401){\makebox(0,0){\strut{}-4}}%
      \csname LTb\endcsname%
      \put(4860,401){\makebox(0,0){\strut{}-2}}%
      \csname LTb\endcsname%
      \put(5308,401){\makebox(0,0){\strut{}0}}%
      \csname LTb\endcsname%
      \put(5755,401){\makebox(0,0){\strut{}2}}%
      \csname LTb\endcsname%
      \put(6203,401){\makebox(0,0){\strut{}4}}%
    }%
    \gplgaddtomacro\gplfronttext{%
      \csname LTb\endcsname%
      \put(4155,1157){\rotatebox{-270}{\makebox(0,0){\strut{}}}}%
      \csname LTb\endcsname%
      \put(5307,196){\makebox(0,0){\strut{}$\wminus / \wfsr$}}%
      \colorrgb{0.06,0.35,0.80}%
      \put(4009,1624){\makebox(0,0){\strut{}\hl{\small{\textbf{(c)}}}}}%
      \colorrgb{0.06,0.35,0.80}%
      \put(5308,1857){\makebox(0,0){\strut{}$\stateAR$}}%
    }%
    \gplgaddtomacro\gplbacktext{%
      \csname LTb\endcsname%
      \put(7042,2047){\makebox(0,0){\strut{}}}%
      \csname LTb\endcsname%
      \put(7490,2047){\makebox(0,0){\strut{}}}%
      \csname LTb\endcsname%
      \put(7938,2047){\makebox(0,0){\strut{}}}%
      \csname LTb\endcsname%
      \put(8385,2047){\makebox(0,0){\strut{}}}%
      \csname LTb\endcsname%
      \put(8833,2047){\makebox(0,0){\strut{}}}%
      \csname LTb\endcsname%
      \put(9154,2244){\makebox(0,0)[l]{\strut{}-1}}%
      \csname LTb\endcsname%
      \put(9154,2804){\makebox(0,0)[l]{\strut{}$0$}}%
      \csname LTb\endcsname%
      \put(9154,3363){\makebox(0,0)[l]{\strut{}1}}%
    }%
    \gplgaddtomacro\gplfronttext{%
      \csname LTb\endcsname%
      \put(6785,2803){\rotatebox{-270}{\makebox(0,0){\strut{}}}}%
      \csname LTb\endcsname%
      \put(7937,1940){\makebox(0,0){\strut{}}}%
      \colorrgb{0.06,0.35,0.80}%
      \put(6639,3270){\makebox(0,0){\strut{}\hl{\small{\textbf{(d)}}}}}%
      \colorrgb{0.06,0.35,0.80}%
      \put(6370,3661){\makebox(0,0){\strut{}JSA}}%
      \colorrgb{0.06,0.35,0.80}%
      \put(7938,3503){\makebox(0,0){\strut{}$\stateRdelayed$}}%
    }%
    \gplgaddtomacro\gplbacktext{%
      \csname LTb\endcsname%
      \put(7042,401){\makebox(0,0){\strut{}-4}}%
      \csname LTb\endcsname%
      \put(7490,401){\makebox(0,0){\strut{}-2}}%
      \csname LTb\endcsname%
      \put(7938,401){\makebox(0,0){\strut{}0}}%
      \csname LTb\endcsname%
      \put(8385,401){\makebox(0,0){\strut{}2}}%
      \csname LTb\endcsname%
      \put(8833,401){\makebox(0,0){\strut{}4}}%
      \csname LTb\endcsname%
      \put(9154,598){\makebox(0,0)[l]{\strut{}-1}}%
      \csname LTb\endcsname%
      \put(9154,1158){\makebox(0,0)[l]{\strut{}$0$}}%
      \csname LTb\endcsname%
      \put(9154,1717){\makebox(0,0)[l]{\strut{}1}}%
    }%
    \gplgaddtomacro\gplfronttext{%
      \csname LTb\endcsname%
      \put(6785,1157){\rotatebox{-270}{\makebox(0,0){\strut{}}}}%
      \csname LTb\endcsname%
      \put(7937,196){\makebox(0,0){\strut{}$\wminus / \wfsr$}}%
      \colorrgb{0.06,0.35,0.80}%
      \put(6639,1624){\makebox(0,0){\strut{}\hl{\small{\textbf{(e)}}}}}%
      \colorrgb{0.06,0.35,0.80}%
      \put(7938,1857){\makebox(0,0){\strut{}$\stateARdelayed$}}%
    }%
    \gplbacktext
    \put(0,0){\includegraphics{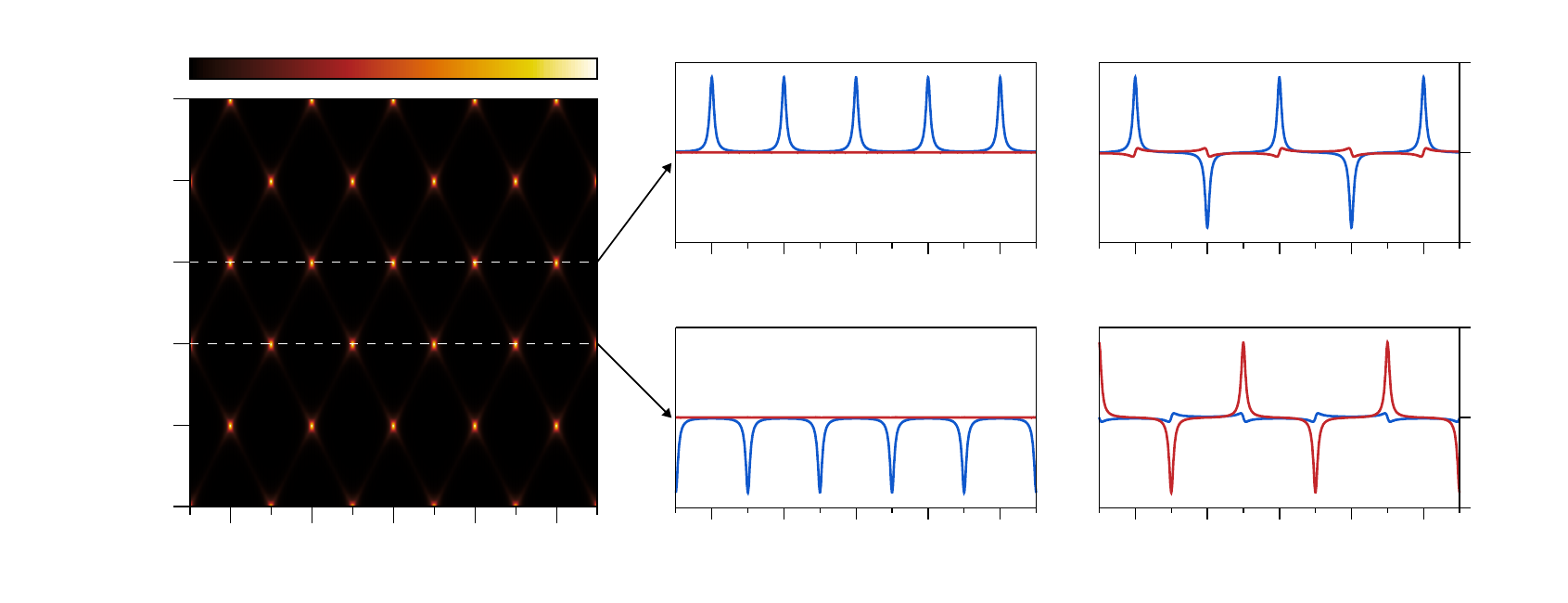}}%
    \gplfronttext
  \end{picture}%
\endgroup

	\caption{a) Simulated JSI of the quantum state generated through type II SPDC filtered by a Fabry Perot resonator with mirror reflectivity 0.8, for a pump beam of linewidth $\Delta\omega \gg \bar{\omega} $ and central frequency $\omega_{0}$, coinciding with the cavity resonance closest to degeneracy. Dashed lines evidence two cuts of the JSI corresponding to a resonant and an anti-resonant state.\\
(b-c) Corresponding simulation of the JSA (blue line: real part; red line: imaginary part), for  $\tau=0$ (d, e) and $\tau=\pi/\bar{\omega}$ (d, e).}
	\label{JSI_Theory}
%\end{multicols}
\end{figure*}

The simulated JSA for intermediate values of the pump beam frequency is presented in the Supplementary Information (See Figure 1 Supplementary material).

In order to control the symmetry of the biphoton state, we introduce a third stage in the experimental setup, consisting of a polarizing beam splitter, separating deterministically the signal and idler photons into two different paths $a$ and $b$, and a delay line on one of the two paths (see Figure \ref{figsetup1}).
The introduced temporal delay $\tau$ modulates the JSA of the state $\ket{\psi}$ with the periodic function $f_{delay}= \exp{(i\tau\wminus)/2}= \cos (\tau\wminus)/2+i\sin(\tau\wminus)/2$, consisting of a symmetric real part and anti-symmetric imaginary part in the $\wminus$ variable. For $\tau= \pi/\bar{\omega}$, corresponding to half of the cavity round trip time, the periodicity $f_{delay}$ is the double of the one of $\ket{\psi}$ state JSA. In this case, the JSA of the resonant state $\ket{\psi_{R}}$ is in phase with the symmetric part of $f_{delay}$, resulting in the pattern of Figure \ref{JSI_Theory}(d). On the contrary the JSA of the anti-resonant state $\ket{\psi_{AR}}$ is in phase with the anti-symmetric part of $f_{delay}$ resulting in the pattern of Figure \ref{JSI_Theory}(e).
The resonant and anti-resonant states after the optical delay stage are:

\begin{equation}
\begin{aligned}
&\ket{\overline{\psi}_{R}}= \sum_{m} C_{PM}(\omega_R,2m\bar{\omega}) e^{im\pi}\\
& \quad\quad\quad\quad\quad \ket{a,\frac{\omega_{R}}{2}+m\bar{\omega}}\ket{b,\frac{\omega_{R}}{2}-m\bar{\omega}} \\
& \ket{\overline{\psi}_{AR}}= \sum_{m} C_{PM}(\omega_{AR},(2m+1)\bar{\omega}) e^{i(m+\frac{1}{2})\pi} \\
& \quad\quad\quad\quad\quad \ket{a,\frac{\omega_{AR}}{2}+(m+\frac{1}{2})\bar{\omega}}\ket{b,\frac{\omega_{AR}}{2}-(m+\frac{1}{2})\bar{\omega}} 
\end{aligned}
\end{equation}  

We note that the state $\ket{\overline{\psi}_{R}}$ is symmetric under particles exchange, while $\ket{\overline{\psi}_{AR}}$ is anti-symmetric.
Analog results occur for $\tau$ values that are odd multiples of the cavity half round-trip time (see Figure 2 in Supplementary Information).

We have thus demonstrated that the proposed experimental setup allows to generate biphoton frequency combs and to control their spectral symmetry by tuning the pump beam frequency. 

In the following, we experimentally demonstrate this method using an AlGaAs chip. This platform combines a large second order optical susceptibility, a direct bandgap and a high electro-optic effect, making it attractive for the miniaturization and the integration of several quantum fonctionalities in a single chip \cite{reviewsara, gunthner2015broadband}. The device consists of a Bragg reflection ridge waveguide optimized for efficient type II SPDC \cite{yeh1976bragg, helmy2006phase, PhysRevLett.112.183901} (See Supplementary Table 1 for details on the structure). % \cite{Valles:13,Weihsprlalggas}; % 
The modes involved in the nonlinear process are a TE Bragg mode for the pump beam around 765 nm and TE$_{00}$ and TM$_{00}$ modes for the photon pairs in the C-telecom band. Note that, for this device, the group velocity mismatch between the two photons of each pair is so small that no off-chip compensation is required to preserve their indistinguishability \cite{Autebertoptica, gunthner2015broadband}. The photon pairs are thus emitted in very good approximation with a joint spectral amplitude centered in $\wminus=0$ and symmetric in the $\wminus$ variable, enabling a direct implementation of our method. Moreover, the refractive index contrast between the semiconductor and the air, leads to a modal reflectivity at the waveguide facets of 0.27(0.24) for the TE(TM) polarized mode, creating a Fabry-Perot cavity surrounding the nonlinear medium. Our chip thus integrates the generation and the cavity stage of the experimental scheme represented in Figure \ref{figsetup1}, leading to an extremely simple and compact solution. Photon pairs are generated by pumping the device with a continuous wave laser having a linewidth of $\Delta\omega = 2\pi \cdot 100 $ kHz, which is much smaller that the free spectral range of the cavity ($\bar{\omega} = 2\pi \cdot 19.2$ GHz).

\begin{figure}[ht]
% GNUPLOT: LaTeX picture with Postscript
\begingroup
\newcommand{\hl}[1]{\setlength{\fboxsep}{0.75pt}\colorbox{white}{#1}}
  \makeatletter
  \providecommand\color[2][]{%
    \GenericError{(gnuplot) \space\space\space\@spaces}{%
      Package color not loaded in conjunction with
      terminal option `colourtext'%
    }{See the gnuplot documentation for explanation.%
    }{Either use 'blacktext' in gnuplot or load the package
      color.sty in LaTeX.}%
    \renewcommand\color[2][]{}%
  }%
  \providecommand\includegraphics[2][]{%
    \GenericError{(gnuplot) \space\space\space\@spaces}{%
      Package graphicx or graphics not loaded%
    }{See the gnuplot documentation for explanation.%
    }{The gnuplot epslatex terminal needs graphicx.sty or graphics.sty.}%
    \renewcommand\includegraphics[2][]{}%
  }%
  \providecommand\rotatebox[2]{#2}%
  \@ifundefined{ifGPcolor}{%
    \newif\ifGPcolor
    \GPcolortrue
  }{}%
  \@ifundefined{ifGPblacktext}{%
    \newif\ifGPblacktext
    \GPblacktexttrue
  }{}%
  % define a \g@addto@macro without @ in the name:
  \let\gplgaddtomacro\g@addto@macro
  % define empty templates for all commands taking text:
  \gdef\gplbacktext{}%
  \gdef\gplfronttext{}%
  \makeatother
  \ifGPblacktext
    % no textcolor at all
    \def\colorrgb#1{}%
    \def\colorgray#1{}%
  \else
    % gray or color?
    \ifGPcolor
      \def\colorrgb#1{\color[rgb]{#1}}%
      \def\colorgray#1{\color[gray]{#1}}%
      \expandafter\def\csname LTw\endcsname{\color{white}}%
      \expandafter\def\csname LTb\endcsname{\color{black}}%
      \expandafter\def\csname LTa\endcsname{\color{black}}%
      \expandafter\def\csname LT0\endcsname{\color[rgb]{1,0,0}}%
      \expandafter\def\csname LT1\endcsname{\color[rgb]{0,1,0}}%
      \expandafter\def\csname LT2\endcsname{\color[rgb]{0,0,1}}%
      \expandafter\def\csname LT3\endcsname{\color[rgb]{1,0,1}}%
      \expandafter\def\csname LT4\endcsname{\color[rgb]{0,1,1}}%
      \expandafter\def\csname LT5\endcsname{\color[rgb]{1,1,0}}%
      \expandafter\def\csname LT6\endcsname{\color[rgb]{0,0,0}}%
      \expandafter\def\csname LT7\endcsname{\color[rgb]{1,0.3,0}}%
      \expandafter\def\csname LT8\endcsname{\color[rgb]{0.5,0.5,0.5}}%
    \else
      % gray
      \def\colorrgb#1{\color{black}}%
      \def\colorgray#1{\color[gray]{#1}}%
      \expandafter\def\csname LTw\endcsname{\color{white}}%
      \expandafter\def\csname LTb\endcsname{\color{black}}%
      \expandafter\def\csname LTa\endcsname{\color{black}}%
      \expandafter\def\csname LT0\endcsname{\color{black}}%
      \expandafter\def\csname LT1\endcsname{\color{black}}%
      \expandafter\def\csname LT2\endcsname{\color{black}}%
      \expandafter\def\csname LT3\endcsname{\color{black}}%
      \expandafter\def\csname LT4\endcsname{\color{black}}%
      \expandafter\def\csname LT5\endcsname{\color{black}}%
      \expandafter\def\csname LT6\endcsname{\color{black}}%
      \expandafter\def\csname LT7\endcsname{\color{black}}%
      \expandafter\def\csname LT8\endcsname{\color{black}}%
    \fi
  \fi
    \setlength{\unitlength}{0.0500bp}%
    \ifx\gptboxheight\undefined%
      \newlength{\gptboxheight}%
      \newlength{\gptboxwidth}%
      \newsavebox{\gptboxtext}%
    \fi%
    \setlength{\fboxrule}{0.5pt}%
    \setlength{\fboxsep}{1pt}%
\begin{picture}(4860.00,2440.00)%
    \gplgaddtomacro\gplbacktext{%
      \csname LTb\endcsname%
      \put(658,895){\makebox(0,0)[r]{\strut{}-1}}%
      \csname LTb\endcsname%
      \put(658,1610){\makebox(0,0)[r]{\strut{}0}}%
      \csname LTb\endcsname%
      \put(1074,377){\makebox(0,0){\strut{}-184}}%
      \csname LTb\endcsname%
      \put(1792,377){\makebox(0,0){\strut{}-182}}%
      \csname LTb\endcsname%
      \put(2510,377){\makebox(0,0){\strut{}-180}}%
      \csname LTb\endcsname%
      \put(3227,377){\makebox(0,0){\strut{}-178}}%
      \csname LTb\endcsname%
      \put(3945,377){\makebox(0,0){\strut{}-176}}%
    }%
    \gplgaddtomacro\gplfronttext{%
      \csname LTb\endcsname%
      \put(274,1280){\rotatebox{-270}{\makebox(0,0){\strut{}$(\wplus - \omega_{R} ) / \wfsr$}}}%
      \csname LTb\endcsname%
      \put(2551,92){\makebox(0,0){\strut{}$\wminus / \wfsr$}}%
      \csname LTb\endcsname%
      \put(2551,2345){\makebox(0,0){\strut{}JSI}}%
      \csname LTb\endcsname%
      \put(729,2264){\makebox(0,0){\strut{}0}}%
      \csname LTb\endcsname%
      \put(4373,2264){\makebox(0,0){\strut{}1}}%
      \colorrgb{0.58,0.00,0.83}%
      \put(4687,1610){\makebox(0,0){\strut{}$\stateR$}}%
      \colorrgb{0.58,0.00,0.83}%
      \put(4737,945){\makebox(0,0){\strut{}$\stateAR$}}%
    }%
    \gplbacktext
    \put(0,0){\includegraphics{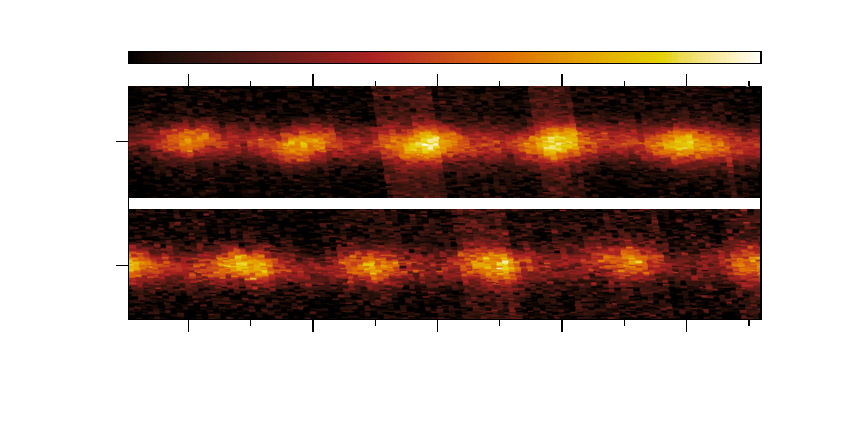}}%
    \gplfronttext
  \end{picture}%
\endgroup

\caption{JSI of the state generated by the AlGaAs chip measured by stimulated emission tomography for two values of the pump beam frequency,  $\omega_p$=$\omega_R$ and $\omega_p$=$\omega_R-\bar{\omega}$ corresponding to a resonant and an anti-resonant state, respectively. $\bar{\omega}=2\pi \cdot 19.2$ GHz, $\omega_R=2\pi \cdot 196.1$ THz (765nm)}
\label{JSI_Measured}
\end{figure}

Figure \ref{JSI_Measured} reports the experimental measurement of the joint spectral intensity JSI=$\lvert C(\omega_+,\omega_-)\rvert^2$ of the quantum state at the output of the AlGaAs chip, measured over a frequency range of 10 $\bar{\omega}$. The measurement is done by stimulated emission tomography \cite{JSIstimulated} for two values of the pump beam frequency separeted by $\bar{\omega}$.
These results prove that the device emits a biphoton frequency comb and that the tuning of the pump frequency controls the transition from an resonant state to an anti-resonant one.

\begin{figure}
% GNUPLOT: LaTeX picture with Postscript
\begingroup
\newcommand{\hl}[1]{\setlength{\fboxsep}{0.75pt}\colorbox{white}{#1}}
  \makeatletter
  \providecommand\color[2][]{%
    \GenericError{(gnuplot) \space\space\space\@spaces}{%
      Package color not loaded in conjunction with
      terminal option `colourtext'%
    }{See the gnuplot documentation for explanation.%
    }{Either use 'blacktext' in gnuplot or load the package
      color.sty in LaTeX.}%
    \renewcommand\color[2][]{}%
  }%
  \providecommand\includegraphics[2][]{%
    \GenericError{(gnuplot) \space\space\space\@spaces}{%
      Package graphicx or graphics not loaded%
    }{See the gnuplot documentation for explanation.%
    }{The gnuplot epslatex terminal needs graphicx.sty or graphics.sty.}%
    \renewcommand\includegraphics[2][]{}%
  }%
  \providecommand\rotatebox[2]{#2}%
  \@ifundefined{ifGPcolor}{%
    \newif\ifGPcolor
    \GPcolortrue
  }{}%
  \@ifundefined{ifGPblacktext}{%
    \newif\ifGPblacktext
    \GPblacktexttrue
  }{}%
  % define a \g@addto@macro without @ in the name:
  \let\gplgaddtomacro\g@addto@macro
  % define empty templates for all commands taking text:
  \gdef\gplbacktext{}%
  \gdef\gplfronttext{}%
  \makeatother
  \ifGPblacktext
    % no textcolor at all
    \def\colorrgb#1{}%
    \def\colorgray#1{}%
  \else
    % gray or color?
    \ifGPcolor
      \def\colorrgb#1{\color[rgb]{#1}}%
      \def\colorgray#1{\color[gray]{#1}}%
      \expandafter\def\csname LTw\endcsname{\color{white}}%
      \expandafter\def\csname LTb\endcsname{\color{black}}%
      \expandafter\def\csname LTa\endcsname{\color{black}}%
      \expandafter\def\csname LT0\endcsname{\color[rgb]{1,0,0}}%
      \expandafter\def\csname LT1\endcsname{\color[rgb]{0,1,0}}%
      \expandafter\def\csname LT2\endcsname{\color[rgb]{0,0,1}}%
      \expandafter\def\csname LT3\endcsname{\color[rgb]{1,0,1}}%
      \expandafter\def\csname LT4\endcsname{\color[rgb]{0,1,1}}%
      \expandafter\def\csname LT5\endcsname{\color[rgb]{1,1,0}}%
      \expandafter\def\csname LT6\endcsname{\color[rgb]{0,0,0}}%
      \expandafter\def\csname LT7\endcsname{\color[rgb]{1,0.3,0}}%
      \expandafter\def\csname LT8\endcsname{\color[rgb]{0.5,0.5,0.5}}%
    \else
      % gray
      \def\colorrgb#1{\color{black}}%
      \def\colorgray#1{\color[gray]{#1}}%
      \expandafter\def\csname LTw\endcsname{\color{white}}%
      \expandafter\def\csname LTb\endcsname{\color{black}}%
      \expandafter\def\csname LTa\endcsname{\color{black}}%
      \expandafter\def\csname LT0\endcsname{\color{black}}%
      \expandafter\def\csname LT1\endcsname{\color{black}}%
      \expandafter\def\csname LT2\endcsname{\color{black}}%
      \expandafter\def\csname LT3\endcsname{\color{black}}%
      \expandafter\def\csname LT4\endcsname{\color{black}}%
      \expandafter\def\csname LT5\endcsname{\color{black}}%
      \expandafter\def\csname LT6\endcsname{\color{black}}%
      \expandafter\def\csname LT7\endcsname{\color{black}}%
      \expandafter\def\csname LT8\endcsname{\color{black}}%
    \fi
  \fi
    \setlength{\unitlength}{0.0500bp}%
    \ifx\gptboxheight\undefined%
      \newlength{\gptboxheight}%
      \newlength{\gptboxwidth}%
      \newsavebox{\gptboxtext}%
    \fi%
    \setlength{\fboxrule}{0.5pt}%
    \setlength{\fboxsep}{1pt}%
\begin{picture}(4860.00,3020.00)%
    \gplgaddtomacro\gplbacktext{%
      \csname LTb\endcsname%
      \put(1170,643){\makebox(0,0)[r]{\strut{}$0$}}%
      \csname LTb\endcsname%
      \put(1170,1367){\makebox(0,0)[r]{\strut{}$0.25$}}%
      \csname LTb\endcsname%
      \put(1170,2091){\makebox(0,0)[r]{\strut{}$0.5$}}%
      \csname LTb\endcsname%
      \put(1170,2815){\makebox(0,0)[r]{\strut{}$0.75$}}%
      \csname LTb\endcsname%
      \put(1343,393){\makebox(0,0){\strut{}$-0.4$}}%
      \csname LTb\endcsname%
      \put(1886,393){\makebox(0,0){\strut{}$-0.2$}}%
      \csname LTb\endcsname%
      \put(2430,393){\makebox(0,0){\strut{}$0$}}%
      \csname LTb\endcsname%
      \put(2973,393){\makebox(0,0){\strut{}$0.2$}}%
      \csname LTb\endcsname%
      \put(3516,393){\makebox(0,0){\strut{}$0.4$}}%
      \csname LTb\endcsname%
      \put(1723,3018){\makebox(0,0)[l]{\strut{}HOM interference}}
      \put(1400,2600){\makebox(0,0)[l]{\strut{}$\ket{{\psi_R}}$}}%%
    }%
    \gplgaddtomacro\gplfronttext{%
      \csname LTb\endcsname%
      \put(570,1729){\rotatebox{-270}{\makebox(0,0){\strut{}Normalized coincidence counts}}}%
      \csname LTb\endcsname%
      \put(2429,125){\makebox(0,0){\strut{}$\tau_{HOM}$ (ps) }}%
    }%
    \gplbacktext
    \put(0,0){\includegraphics{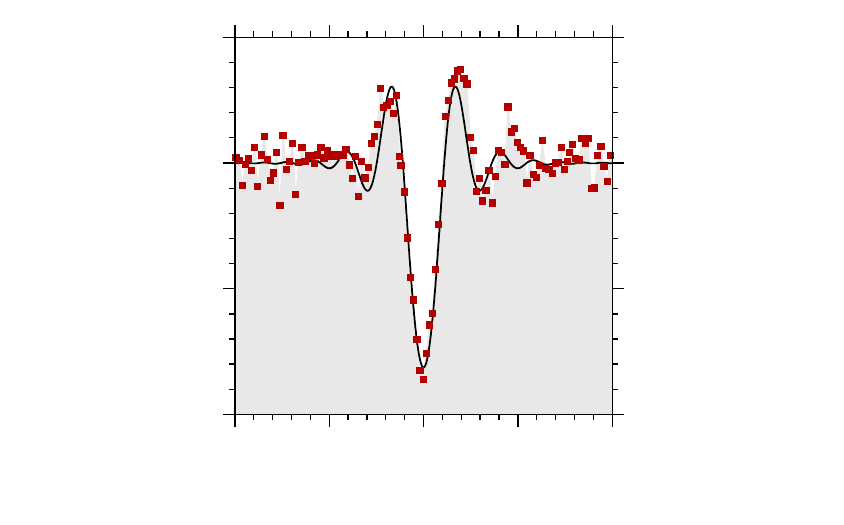}}%
    \gplfronttext
  \end{picture}%
\endgroup

\caption{Results of the HOM measurement of the state $\ket{{\psi_R}}$ produced at the output of the AlGaAs chip. Squares: experimental data. Line: theoretical model.The obtained visibility is 0.86 and the spectral bandwidth of the interfering photons is 170 nm.}
\label{HOM_measured_center}
\end{figure}

In order to quantify the level of mirror symmetry of the JSA function, we implement a HOM interferometer. Generated photons are separated with a fibered polarizing beam splitter, delayed by $\tau_{HOM}$, their polarisation aligned with a half wave-plate, and they are recombined at a fibered 50/50 beam splitter. The two output ports are connected to single photon avalanche photodiodes having a detection efficiency of 25$\%$ and the coincidences counts are recorded with a time-to-digital converter. The obtained results for $\ket{\psi_{R}}$ are reported in Figure \ref{HOM_measured_center}: a dip having a width of 52 $\pm 2 $ fs is observed. Similar results are obtained for $\ket{\psi_{AR}}$. Its  visibility defined as $(N_{\tau}-N_{0})/N_{\tau})$, where $N_{\tau}$ is the coincidences rate far from the interference region and $N_{0}$ the coincidence rate at the minimum of the dip, is 86$\%$. This value is limited by residual modal birefringence. The oscillating behavior observed around the dip is well-described by taking into account the chromatic dispersion of the sample, as shown by the result of the numerical simulation reported in Figure \ref{HOM_measured_center}.
From this simulation we can extract the emission bandwidth of the photon pairs, $\Delta\wminus = 2\pi \cdot 21.82$ THz, corresponding to a signal and idler bandwidth of $\Delta\lambda_{s,i} = 170.9$ nm around 1530 nm (see details in Supplementary Information Figure 3-4). This result demonstrates that the device generates a broadband biphoton frequency comb, where the photons of each pair are in a coherent superposition of more than 500 peaks.

The control of the symmetry of the generated state is done by implementing the last stage of the experimental scheme presented in Figure \ref{figsetup1}, that is by adding a delay line on one of the output ports of the polarizing beam splitter and by choosing a temporal delay $\tau= \pi/\bar{\omega}$. This leads to the generation of the state $\ket{\overline{\psi}_{R}}$ for the resonant case and $\ket{\overline{\psi}_{AR}}$ for the anti-resonant case.

\begin{figure}
% GNUPLOT: LaTeX picture with Postscript
\begingroup
\newcommand{\hl}[1]{\setlength{\fboxsep}{0.75pt}\colorbox{white}{#1}}
  \makeatletter
  \providecommand\color[2][]{%
    \GenericError{(gnuplot) \space\space\space\@spaces}{%
      Package color not loaded in conjunction with
      terminal option `colourtext'%
    }{See the gnuplot documentation for explanation.%
    }{Either use 'blacktext' in gnuplot or load the package
      color.sty in LaTeX.}%
    \renewcommand\color[2][]{}%
  }%
  \providecommand\includegraphics[2][]{%
    \GenericError{(gnuplot) \space\space\space\@spaces}{%
      Package graphicx or graphics not loaded%
    }{See the gnuplot documentation for explanation.%
    }{The gnuplot epslatex terminal needs graphicx.sty or graphics.sty.}%
    \renewcommand\includegraphics[2][]{}%
  }%
  \providecommand\rotatebox[2]{#2}%
  \@ifundefined{ifGPcolor}{%
    \newif\ifGPcolor
    \GPcolortrue
  }{}%
  \@ifundefined{ifGPblacktext}{%
    \newif\ifGPblacktext
    \GPblacktexttrue
  }{}%
  % define a \g@addto@macro without @ in the name:
  \let\gplgaddtomacro\g@addto@macro
  % define empty templates for all commands taking text:
  \gdef\gplbacktext{}%
  \gdef\gplfronttext{}%
  \makeatother
  \ifGPblacktext
    % no textcolor at all
    \def\colorrgb#1{}%
    \def\colorgray#1{}%
  \else
    % gray or color?
    \ifGPcolor
      \def\colorrgb#1{\color[rgb]{#1}}%
      \def\colorgray#1{\color[gray]{#1}}%
      \expandafter\def\csname LTw\endcsname{\color{white}}%
      \expandafter\def\csname LTb\endcsname{\color{black}}%
      \expandafter\def\csname LTa\endcsname{\color{black}}%
      \expandafter\def\csname LT0\endcsname{\color[rgb]{1,0,0}}%
      \expandafter\def\csname LT1\endcsname{\color[rgb]{0,1,0}}%
      \expandafter\def\csname LT2\endcsname{\color[rgb]{0,0,1}}%
      \expandafter\def\csname LT3\endcsname{\color[rgb]{1,0,1}}%
      \expandafter\def\csname LT4\endcsname{\color[rgb]{0,1,1}}%
      \expandafter\def\csname LT5\endcsname{\color[rgb]{1,1,0}}%
      \expandafter\def\csname LT6\endcsname{\color[rgb]{0,0,0}}%
      \expandafter\def\csname LT7\endcsname{\color[rgb]{1,0.3,0}}%
      \expandafter\def\csname LT8\endcsname{\color[rgb]{0.5,0.5,0.5}}%
    \else
      % gray
      \def\colorrgb#1{\color{black}}%
      \def\colorgray#1{\color[gray]{#1}}%
      \expandafter\def\csname LTw\endcsname{\color{white}}%
      \expandafter\def\csname LTb\endcsname{\color{black}}%
      \expandafter\def\csname LTa\endcsname{\color{black}}%
      \expandafter\def\csname LT0\endcsname{\color{black}}%
      \expandafter\def\csname LT1\endcsname{\color{black}}%
      \expandafter\def\csname LT2\endcsname{\color{black}}%
      \expandafter\def\csname LT3\endcsname{\color{black}}%
      \expandafter\def\csname LT4\endcsname{\color{black}}%
      \expandafter\def\csname LT5\endcsname{\color{black}}%
      \expandafter\def\csname LT6\endcsname{\color{black}}%
      \expandafter\def\csname LT7\endcsname{\color{black}}%
      \expandafter\def\csname LT8\endcsname{\color{black}}%
    \fi
  \fi
    \setlength{\unitlength}{0.0500bp}%
    \ifx\gptboxheight\undefined%
      \newlength{\gptboxheight}%
      \newlength{\gptboxwidth}%
      \newsavebox{\gptboxtext}%
    \fi%
    \setlength{\fboxrule}{0.5pt}%
    \setlength{\fboxsep}{1pt}%
\begin{picture}(4860.00,3020.00)%
    \gplgaddtomacro\gplbacktext{%
      \csname LTb\endcsname%
      \put(556,604){\makebox(0,0)[r]{\strut{}$0.4$}}%
      \csname LTb\endcsname%
      \put(556,1551){\makebox(0,0)[r]{\strut{}$0.5$}}%
      \csname LTb\endcsname%
      \put(556,2498){\makebox(0,0)[r]{\strut{}$0.6$}}%
      \csname LTb\endcsname%
      \put(1102,354){\makebox(0,0){\strut{}$-0.2$}}%
      \csname LTb\endcsname%
      \put(1676,354){\makebox(0,0){\strut{}$0$}}%
      \csname LTb\endcsname%
      \put(2250,354){\makebox(0,0){\strut{}$0.2$}}%
      \csname LTb\endcsname%
      \put(815,2309){\makebox(0,0)[l]{\strut{}$\stateRdelayed$}}%
    }%
    \gplgaddtomacro\gplfronttext{%
      \csname LTb\endcsname%
      \put(58,1551){\rotatebox{-270}{\makebox(0,0){\strut{}Normalized coincidence counts}}}%
      \csname LTb\endcsname%
      \put(1676,86){\makebox(0,0){\strut{}$\tau_{HOM}$ (ps) }}%
    }%
    \gplgaddtomacro\gplbacktext{%
      \csname LTb\endcsname%
      \put(2694,604){\makebox(0,0)[r]{\strut{}}}%
      \csname LTb\endcsname%
      \put(2694,1551){\makebox(0,0)[r]{\strut{}}}%
      \csname LTb\endcsname%
      \put(2694,2498){\makebox(0,0)[r]{\strut{}}}%
      \csname LTb\endcsname%
      \put(3240,354){\makebox(0,0){\strut{}$-0.2$}}%
      \csname LTb\endcsname%
      \put(3814,354){\makebox(0,0){\strut{}$0$}}%
      \csname LTb\endcsname%
      \put(4388,354){\makebox(0,0){\strut{}$0.2$}}%
      \csname LTb\endcsname%
      \put(2953,2309){\makebox(0,0)[l]{\strut{}$\stateARdelayed$}}%
      \csname LTb\endcsname%
      \put(1977,2782){\makebox(0,0)[l]{\strut{}HOM interference}}%
    }%
    \gplgaddtomacro\gplfronttext{%
      \csname LTb\endcsname%
      \put(3814,86){\makebox(0,0){\strut{}$\tau_{HOM}$ (ps) }}%
    }%
    \gplbacktext
    \put(0,0){\includegraphics{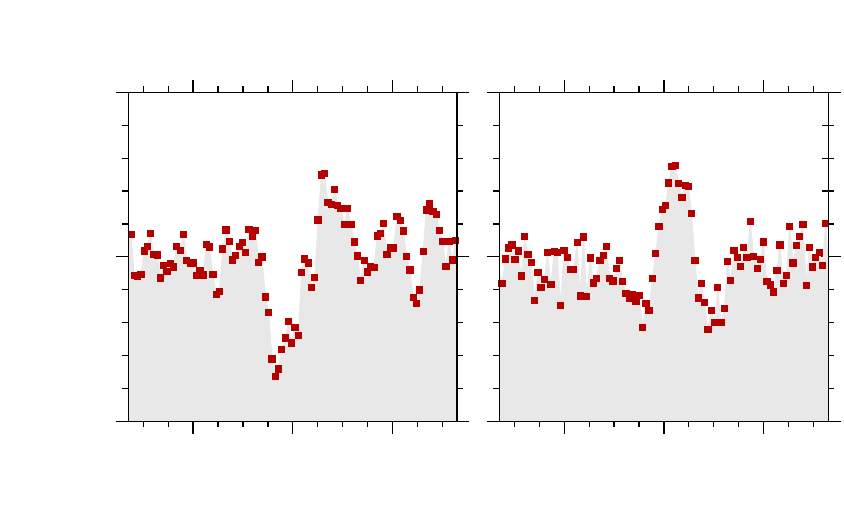}}%
    \gplfronttext
  \end{picture}%
\endgroup

\caption{Results of the HOM measurement of the (left) resonant state $\ket{\overline{\psi}^{R}}$ and (right) anti-resonant state $\ket{\overline{\psi}^{AR}}$. }
\label{HOM_measured_replica}
\end{figure}

In order to evaluate the symmetry of the JSA of these states, we rely on HOM interferometry: photon pairs described by a symmetric JSA are expected to bunch, whereas photon pairs described by an anti-symmetric wave function are expected to anti-bunch \cite{fedrizzi2009anti}. Figure \ref{HOM_measured_replica} reports the experimental results of the HOM experiment: when we inject in the beam splitter the state $\ket{\overline{\psi}_{R}}$ a dip is observed, while when we inject in the beam splitter the state $\ket{\overline{\psi}_{AR}}$ we obtain a peak. The observed visibility is $\approx 10 \% $, due to the combined effect of reflectivity \cite{Sagioro2004}, birefringence and chromatic dispersion. Taking into account these effects, we predict that a spectral filter, centered at the frequency degeneracy and having a bandwidth of 25 nm, together with a reflection coating increasing the facets reflectivity to 0.5, would allow to reach a visibility of $70 \%$.
We discuss the estimated visibility in the ideal case of zero birefringence (or compensated birefringence) and zero chromatic dispersion as a function of the reflectivity in the Supplementary material Figure 5.

We underline that the use of frequency anti-correlations and of a pump beam with a narrow spectral profile ($\Delta\omega << \bar{\omega}$) are essential elements for the manipulation of the state symmetry; a broad pump would generate an uncorrelated state, consisting of a superposition of resonant and anti-resonant patterns, incompatible with this symmetry manipulation protocol.

In conclusion, we have proposed and demonstrated a method to generate and manipulate the symmetry of biphoton frequency combs based on the interplay between cavity effects and a delay line. The method can be adapted and applied to a large variety of systems, either bulk or integrated, thus increasing their flexibility and the richness of the generated states. In addition since it doesn't rely on post selection, it can generate on demand anti-symmetric states. 
We have shown that AlGaAs Bragg reflector waveguides are a particularly convenient system to implement this method, thanks to the emission of photon pairs via type II SPDC (leading to a deterministic separation of the photon pairs), small birefringence of the generated modes (leading to a symmetric JSA with respect to frequency degeneracy and avoiding the requirement of off-chip compensation), and to the natural presence of a cavity (due to facets reflectivity). Further progress is possible in different ways: a full integration of the setup could be obtained by developing an on-chip polarizing beam splitter, as already demonstrated in Si-based devices \cite{cai2017silicon}, and by controlling the relative delay between the orthogonally polarized photons through the electro-optic effect \cite{wang2014gallium}. Moreover, the compliance AlGaAs chip with electrical injection at room temperature \cite{PhysRevLett.112.183901} paves the way towards the integration of the laser source within the chip, resulting in an extremely miniaturized and versatile system.  
These results open the way to new quantum protocols exploiting high-dimensional frequency states with controllable symmetry, such as the implementation of quantum logic gates through coherent manipulation of entangled frequency-bin qubits \cite{jaramillo2017persistent}, high-dimensional one-way quantum processing \cite{reimer2018high} or error correction in high-dimensional redundant states \cite{GKP2019inpreparation}.

The authors gratefully acknowledge ANR (Agence Nationale de la Recherche) for the financial support of this work through Project SemiQuantRoom (Project No. ANR-14-CE26-0029) and through Labex SEAM (Science and Engineering for Advanced Materials and devices) project ANR 11 LABX 086, ANR 11 IDEX 05 02.The French RENATECH network and Universit\'e Sorbonne Paris Cit\'e for PhD fellowship to G.M. are also warmly acknowledged.

\bibliographystyle{apsrev4-1}
\bibliography{bibliography}

\end{document}